\documentclass[preprint,12pt]{elsarticle}

\usepackage{epsfig}
\usepackage{enumerate}

\usepackage{amssymb,amsmath,graphics,graphicx}  
\usepackage{xcolor}

\journal{Elsevier}

\begin{document}

\begin{frontmatter}


\title{Modeling the evolution of drinking behavior: A Statistical Physics perspective}



\author{Nuno Crokidakis}
\ead{nuno@mail.if.uff.br}
\author{Lucas Sigaud}

\address{
Instituto de F\'{\i}sica, Universidade Federal Fluminense, Niter\'oi, Rio de Janeiro, Brazil
}

\begin{abstract}
In this work we study a simple compartmental model for drinking behavior evolution. The population is divided in 3 compartments regarding their alcohol consumption, namely Susceptible individuals $S$ (nonconsumers), Moderate drinkers $M$ and Risk drinkers $R$. The transitions among those states are ruled by probabilities. Despite the simplicity of the model, we observed the occurrence of two distinct nonequilibrium phase transitions to absorbing states. One of these states is composed only by Susceptible individuals $S$, with no drinkers ($M=R=0$). On the other hand, the other absorbing state is composed only by Risk drinkers $R$ ($S=M=0$). Between these two steady states, we have the coexistence of the three subpopulations $S$, $M$ and $R$. Comparison with abusive alcohol consumption data for Brazil shows a good agreement between the model's results and the database.

\end{abstract}


\end{frontmatter}

%\linenumbers

\section{Introduction}
\label{S:1}

Epidemic models have been widely used to study contagion processes such as the spread of infectious diseases \cite{bailey1975mathematical} and rumors \cite{daley1964epidemics}. This kind of model has also been used for the spread of social habits, such as the smoking habit \cite{PMID:21696936}, cocaine \cite{sanchez2011predicting} and alcohol consumption \cite{santonja2010alcohol}, obesity \cite{ejima2013modeling}, corruption \cite{crokidakis2018can}, cooperation \cite{lima2014evolution}, ideological conflicts \cite{marvel2012encouraging}, and also to other problems like rise/fall of ancient empires \cite{gunduz2016dynamics}, dynamics of tax evasion \cite{brum2017dynamics}, radicalization phenomena \cite{galam2016modeling}, fanaticism \cite{stauffer2007can} and violence \cite{nizamani2014public}.

The main reason such social behaviors can be modelled by contagion processes is the response by elements of the ensemble to the social context of the studied subject. Both social or peer pressure and positive reinforcement from other agents, regardless if the behavior brings positive or negative consequences to the individual, can influence each one's way of life. Therefore, models for the epidemics of infectious diseases are also able to describe the spread of such tendencies, like alcoholism \cite{galea2009social,gorman2006agent}.

The standard medical way of categorizing alcohol consumption \cite{oms2018} is in three groups - nonconsumers, moderate (or social) consumers and risk (or excessive) consumers; thus, modeling of the interactions and consequent changes of an individual from one group to another is governed by interaction parameters. One interesting aspect that should be taken into consideration when modeling alcohol consumption is the tendency on some individuals to gradually increase their consumption rate, not due to social susceptibility, but when under stressful or depressing circumstances, since alcohol plays a major role both as cause and consequence of depression, for instance \cite{sullivan2005}. This means that one can attach a probability of a moderate drinker to become an excessive drinker that is dependent only on the actual moderate drinkers population size, instead of the two population groups involved in the change. If one considers the current world situation with the recent coronavirus disease 2019 outbreak (COVID-19), this self-induced increase in alcohol consumption is not only realistic, but also becomes more prominent - this has been observed in a myriad of studies this year detailing the consequences and dangers of both alcohol withdrawal (in places where it has become harder to legally acquire alcohol during the pandemic) and alcohol consumption increase \cite{chodkiewicz2020alcohol,clay2020alcohol,narasimha2020complicated,rehm2020alcohol}.

This work is organized as follows. In Section 2, we present the overall panorama of different epidemic compartment models for drinking behavior as well as this work's model and define the microscopic rules that lead to its dynamics. The analytical and numerical results are presented in Section 3, including comparisons with Brazil's alcohol consumption data for a range of eleven years, used as a case study in order to evaluate the present model. Finally, our conclusions are presented in section 4.

\newpage

% ##############################################################

\section{Model}

Our model is based on the proposal of references \cite{santonja2010alcohol,galea2009social,gorman2006agent,walters2013modelling,nazir2019conformable,huo2018dynamics,mulone2012modeling,sanchez2007drinking,sharma2013drinking,agrawal2018role,adu2017mathematical,huo2012global,khajji2020discrete,huo2017optimal,muthuri2019modeling,ma2015modelling,wang2014optimal} that treat alcohol consumption as a disease that spreads by social interactions and a brief outlook of many works in the area will be presented in the following.

Santonja \textit{et al.} have proposed a three-compartment model divided in nonconsumers, nonrisk consumers and risk consumers, with only one social interaction - drinkers influence nonconsumers to become nonrisk consumers; other interactions (from nonrisk to risk consumers and from risk to non-consumers) happen spontaneously \cite{santonja2010alcohol}. Gorman \textit{et al.}, on the other hand, treats the three-compartment model as composed of susceptible drinkers, current drinkers and former drinkers (a definitive compartment), separating the nonconsumer population into two compartments, while concentrating all consumers in a single one \cite{gorman2006agent}. Walters \textit{et al.} propose a similar model to Gorman \textit{et al.}, but treating the former drinkers compartment as an in-treatment compartment, from which individuals can relapse and, thus, repopulate the drinkers compartment \cite{walters2013modelling}. 

Evolving as a combination of these two previously described works, Sharma and Samanta proposed a four-compartment model, combining Walters \textit{et al.} model with the addition of a recovered group, which plays the same role as the former drinkers compartment in Gorman \textit{et al.} model with the exception that it is treated as temporary, with individuals returning to the non-consumers compartment (but never returning to in-treatment or consumers compartment directly) \cite{sharma2013drinking}. Agrawal, Tenguria and Modi \cite{agrawal2018role} further develop this modelling structure by incorporating nonlinear incidence rates explored previously by Adu \textit{et al.} \cite{adu2017mathematical}. In these works, both endemic and drinking-free equilibrium states were discussed. Huo and Song \cite{huo2012global} included a compartmental distinction between heavy drinkers who admit or not to drinking, incorporating a two-stage process from susceptible to admitting heavy drinkers. More recently, Khajji \textit{et al.} proposed a five-compartment model in which individuals can migrate from susceptible to moderate to heavy, to in-treatment, to recovered, further subdividing the in-treatment compartment into populations in public treatment centers and private treatment centers in order to evaluate how economic differences influence recovery rates \cite{khajji2020discrete}.

Many different works have also dedicated themselves to try to model external influences on recovery of risk consumers. For example, Huo \textit{et al.} \cite{huo2017optimal} and Muthuri \textit{et al.} \cite{muthuri2019modeling} have attempted to quantitatively incorporate the effect of mass media campaigns in their models. Awareness programs have originated time delay shifts on the progression of compartment mobility on Ma \textit{et al.} model \cite{ma2015modelling}, while Wang \textit{et al.} \cite{wang2014optimal} have investigated nonlinear approaches to such time delay shift of hindered interactions between susceptible and consumer individuals.

With such context in mind, the presented model aims, in a simple three-compartment model, to analyze the influence of social interaction by keeping all possible mobility between compartments dependent on populations on both departure and arrival compartments, with the exception of a spontaneous - and competing with the social - probability of moderate to risk consumer transition. In such case, we consider an epidemic-like model where the transitions among the compartments are governed by probabilities. In this work we consider population homogeneous mixing, i.e., a fully-connected population of $N$ individuals. This population is divided in 3 compartments, namely:

\begin{itemize}
  
\item \textbf{S}: nonconsumer individuals, individuals that have never consumed alcohol or have consumed in the past and quit. In this case, we will call them Susceptible individuals, i.e., susceptible to become drinkers, either again or for the first time;

\item \textbf{M}: nonrisk consumers, individuals with regular low consumption. We will call them Moderate drinkers;

\item \textbf{R}: risk consumers, individuals with regular high consumption. We will call them Risk drinkers;
\end{itemize}

To be precise, a moderate drinker is a man who consumes less than 50 cc of alcohol every day or a woman who consumes less than 30 cc of alcohol every day. On the other hand, a risk drinker is a man who consumes more than 50 cc of alcohol every day or a woman who consumes more than 30 cc of alcohol every day \cite{santonja2010alcohol} \footnote{Alternatively, Brazil's Ministry of Health also includes people with frequent episodes of excessive drinking - namely 100 cc for men and 80 cc for women in one occasion - to be Risk drinkers \cite{databrasil}.}. Since we are considering a contagion model, the probabilities related to changes in agents' compartments represent the possible contagions. The transitions among compartments are as following:

\begin{itemize}
\item $S \stackrel{\beta}{\rightarrow} M$: a Susceptible agent (\textbf{S}) becomes a Moderate drinker (\textbf{M}) with probability $\beta$ if he/she is in contact with Moderate (\textbf{M}) or Risk (\textbf{R}) drinkers;

\item $M \stackrel{\alpha}{\rightarrow} R$: a Moderate drinker (\textbf{M}) becomes a Risk drinker (\textbf{R}) with probability $\alpha$ ;

\item $M \stackrel{\delta}{\rightarrow} R$: a Moderate drinker (\textbf{M}) becomes a Risk drinker (\textbf{R}) with probability $\delta$ if he/she is in contact with Risks drinkers (\textbf{R});

\item $R \stackrel{\gamma}{\rightarrow} S$: a Risk drinker (\textbf{R}) becomes a Susceptible agent (\textbf{S}) with probability $\gamma$ if he/she is in contact with Susceptible individuals (\textbf{S});  

\end{itemize}

In the above rules, $\beta$ represents an ``infection'' probability, i.e., the probability that a consumer (M or R) individual turns a nonconsumer one into drinker. The Risk drinkers R can also ``infect'' the Moderate M agents and turn them into Risk drinkers R, which occurs with probability $\delta$. These two infections occur by contagion, in our model, where individuals belonging to a group with a higher degree of consumption can influence others to drink more via social contact. This transition $M\to R$ can also occur spontaneously, with probability $\alpha$, if a given agent increase his/her alcohol consumption - this is the only migration pathway from one group to another, in this model, that does not depend on the population of the receiving compartment, since it corresponds to a self-induced progression from Moderate (\textbf{M}) to Risk (\textbf{R}) drinking. As stated in the introduction, above, the increase of alcohol consumption has been documented to occur under stressful circumstances (like the COVID-19 pandemic) or clinical depression, regardless of social interaction with Risk drinkers. Finally, the probability $\gamma$ represents the infection probability that turn Risk drinkers $R$ into Susceptible agents $S$. In this case, it can represent the pressure of social contacts (family, friends, etc) over individuals that drink excessively. We did not take into account transitions from Risk (\textbf{R}) to Moderate (\textbf{M}), assuming that, as a rule, once an individual reaches a behavior of excessive consumption of alcohol, contact with Moderate drinkers does not imply on a tendency to lower one's consumption - meanwhile, it is assumed that contacts that do not drink at all are able to exert a higher pressure on them to quit drinking. It is not that the from Risk to Moderate transition cannot occur - it is just that for our model this probability, when comparing it with the overall picture, is negligible.

For simplicity, we consider a fixed population, i.e., at each time step $t$ we have the normalization condition $s(t)+m(t)+r(t)=1$, where we defined the population densities $s(t)=S(t)/N$, $m(t)=M(t)/N$ and $r(t)=R(t)/N$. Since we will only deal with the relative proportions among the three different groups in relation to the total population $N$, i.e. the population densities, we will not take into account birth-mortality relations and populational increase/decrease effects. So, even if $N$ is not a constant number, for all modelling purposes it will not matter due to the fact that we will deal only with the $s(t)$, $m(t)$ and $r(t)$ subpopulations in relation to the total population. One other way of looking at this approximation is to consider only the adult population as relevant to our modelling, and assume that new individuals coming of age correspond to the number of deaths \cite{mulone2012modeling,sanchez2007drinking}.

% ###############################################################

\section{Results}

Based on the microscopic rules defined in the previous subsection, one can write the master equations that describe the time evolution of the densities $s(t)$, $m(t)$ and $r(t)$ as follows,
\begin{eqnarray} \label{eq1}
\frac{ds(t)}{dt} & = & -\beta\,s(t)\,m(t) - \beta\,s(t)\,r(t) + \gamma\,s(t)\,r(t) ~, \\ \label{eq2}
\frac{dm(t)}{dt} & = & \beta\,s(t)\,m(t) + \beta\,s(t)\,r(t) - \delta\,m(t)\,r(t) - \alpha\,m(t) ~, \\ \label{eq3}
\frac{dr(t)}{dt} & = & \alpha\,m(t) + \delta\,m(t)\,r(t) - \gamma\,s(t)\,r(t) ~,
\end{eqnarray}  
and we also have the normalization condition
\begin{equation} \label{eq4}
s(t)+m(t)+r(t)=1 ~,
\end{equation}
valid at each time step $t$. 

First of all, one can analyze the early evolution of the population, for small times. As it is usual in epidemic models, we consider as initial condition the introduction of a single infected individual. In our case, one moderate drinker, i.e., our initial conditions are given by the densities $m(0)=1/N$, $s(0)=1 - 1/N$ and $r(0)=0$. In such case, one can linearize Eq. (\ref{eq2}) to obtain \cite{bailey1975mathematical}
\begin{equation} \label{eq5}
\frac{dm(t)}{dt}=(\beta-\alpha)\,m(t) ~,
\end{equation}
\noindent
that can be directly integrated to obtain $m(t)=m_0\,e^{\alpha(R_0-1)\,t}$, where $m_0=m(t=0)$ and one can obtain the expression for the basic reproduction number
\begin{equation} \label{eq6}
R_0 = \frac{\beta}{\alpha} ~.
\end{equation}
As it is usual in epidemic models \cite{bailey1975mathematical,crokidakis2020covid}, the disease (alcoholism) will persist in the population if $R_0>1$, i.e., for $\beta>\alpha$.

%%%%%%%%%%%%%%%%%%%%%%%%%%%%%%%%%%%%%%%%%%%%%%%%%%%%%%%%%%%%%%%%
\begin{figure}[t]
\begin{center}
\vspace{6mm}
\includegraphics[width=0.45\textwidth,angle=0]{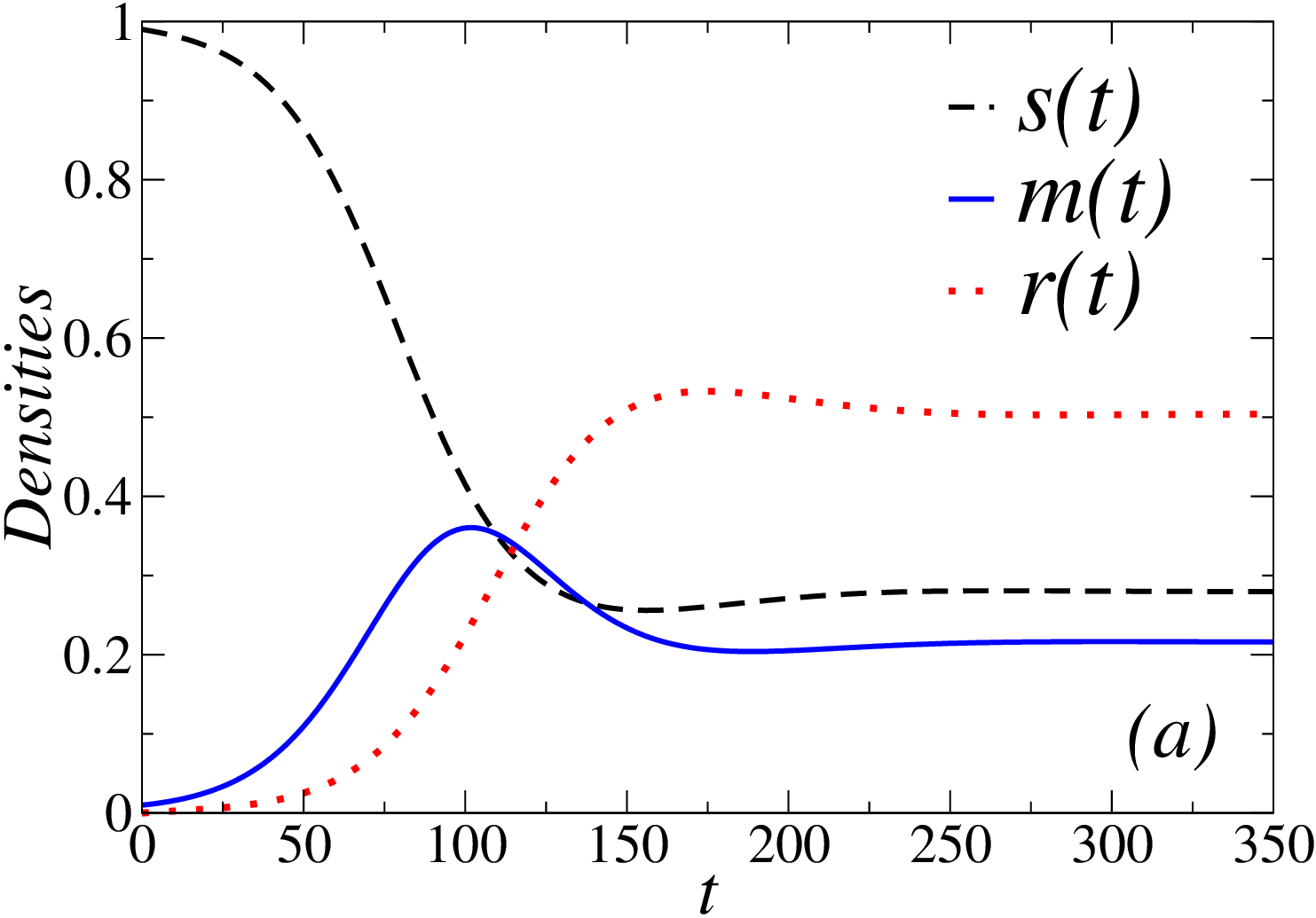}
\hspace{0.3cm}
\includegraphics[width=0.45\textwidth,angle=0]{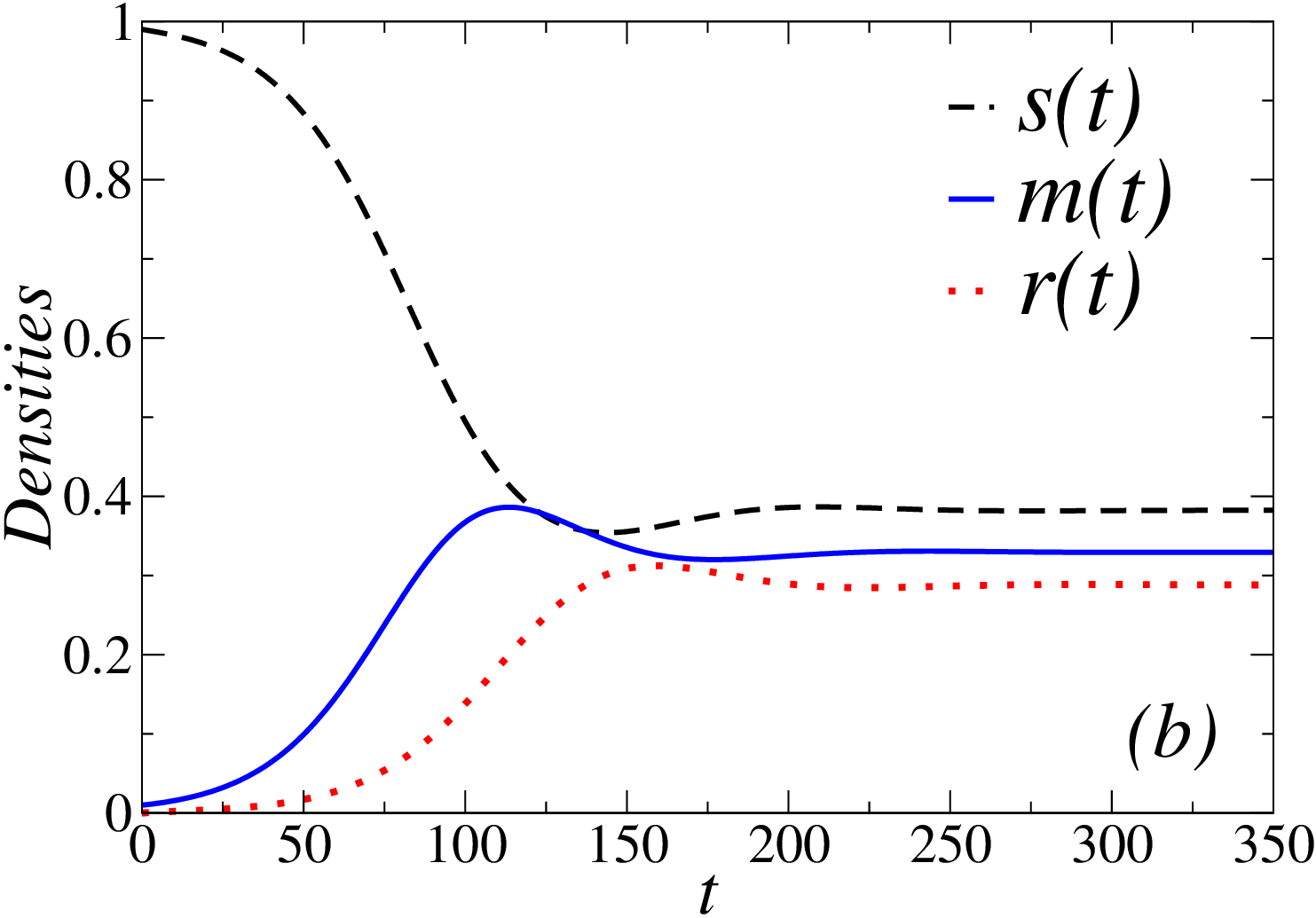}
\\
\vspace{0.5cm}
\includegraphics[width=0.45\textwidth,angle=0]{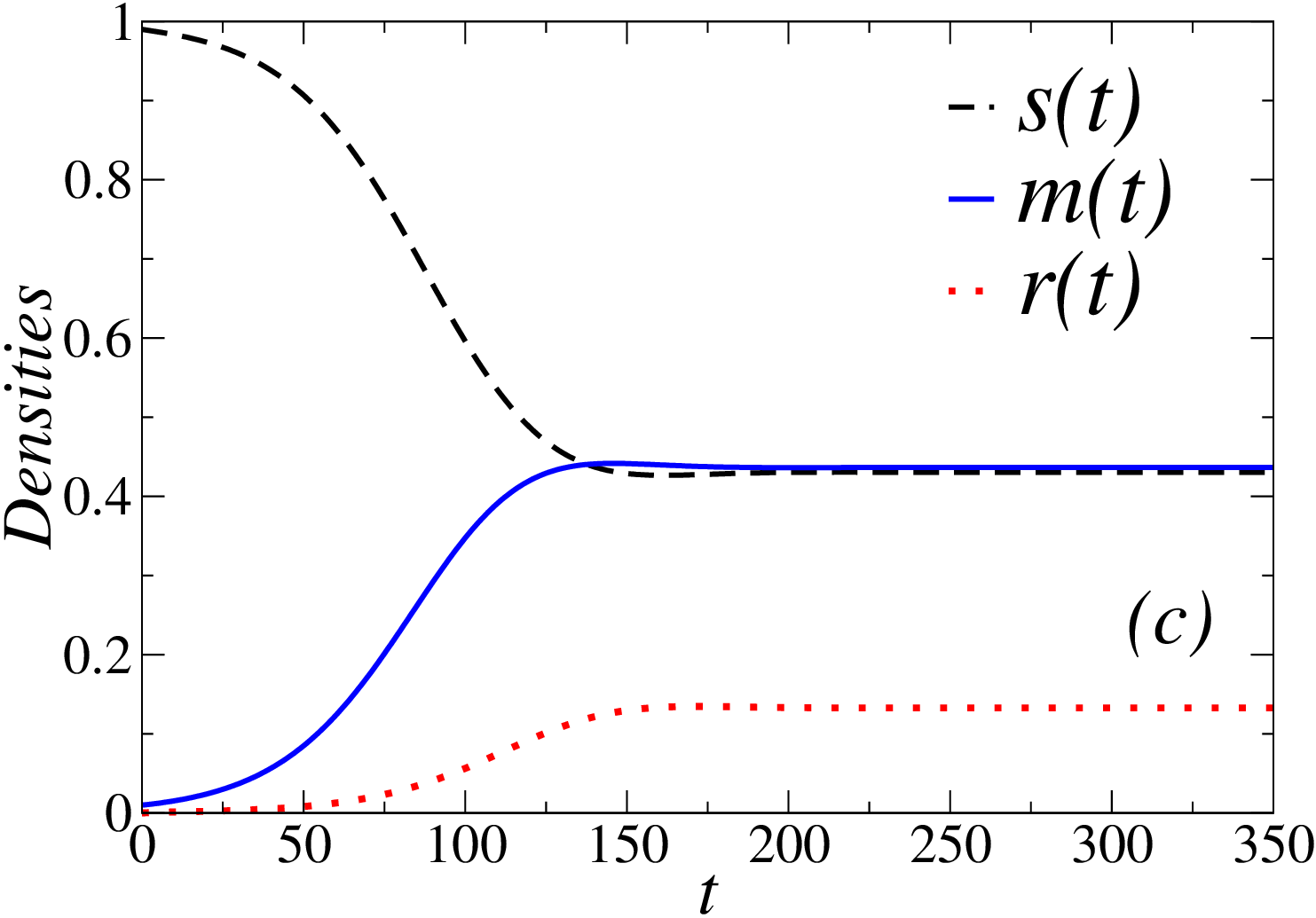}
\hspace{0.3cm}
\includegraphics[width=0.45\textwidth,angle=0]{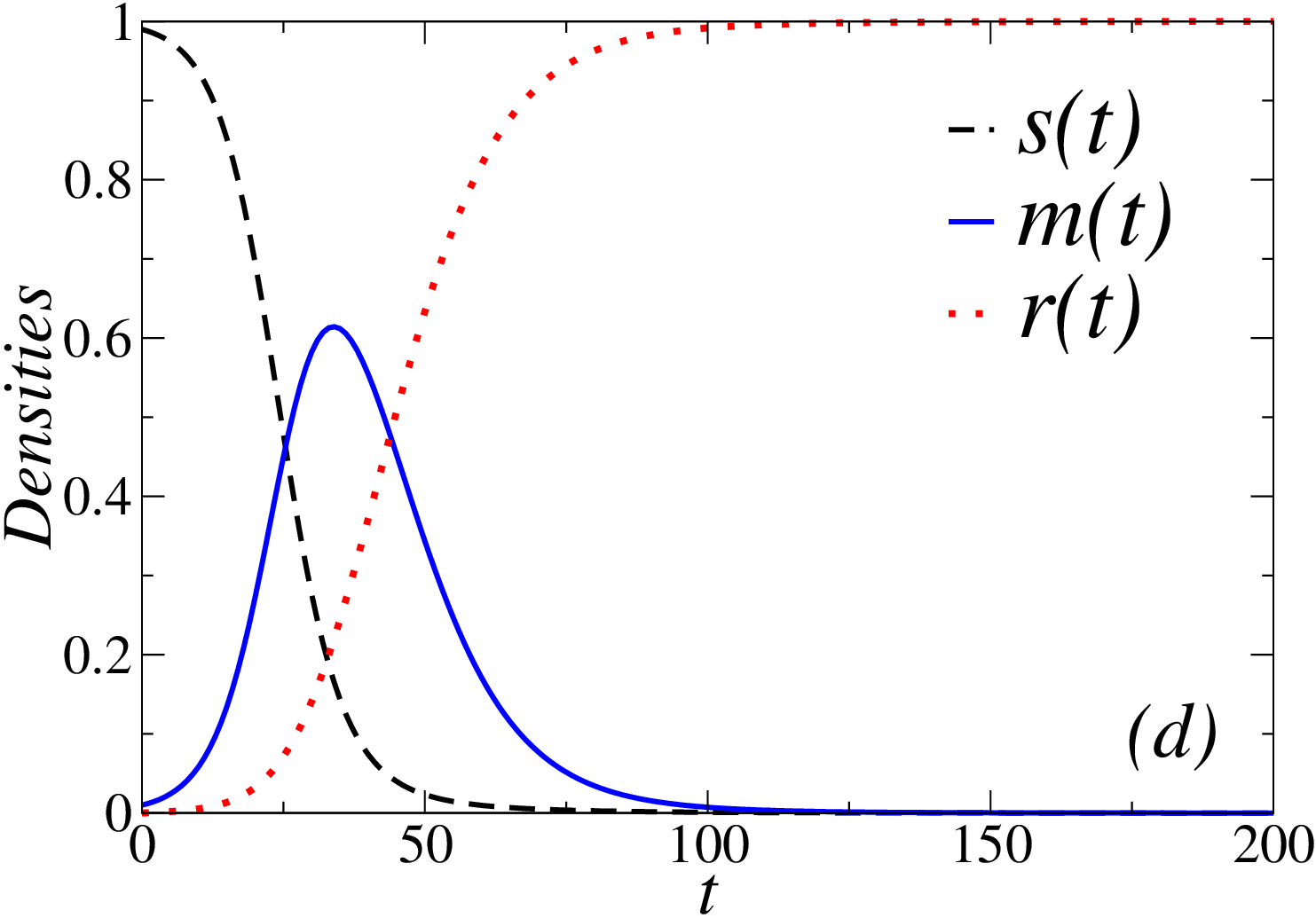}

\end{center}
\caption{(Color online) Time evolution of the three densities of agents $s(t)$, $m(t)$ and $r(t)$, based on the numerical integration of Eqs. (\ref{eq1}) - (\ref{eq3}). The fixed parameters are $\alpha=0.03$ and $\delta=0.07$, and we varied the parameters $\beta$ and $\gamma$: (a) $\beta=0.07, \gamma=0.10$, (b) $\beta=0.07, \gamma=0.15$, (c) $\beta=0.07, \gamma=0.30$, (d) $\beta=0.20, \gamma=0.15$. From Eq. (\ref{eq6}), we obtain $R_0\approx 2.33$ for the panels (a)-(c) and $R_0\approx 6.67$ for panel (d).}
\label{fig1}
\end{figure}
%%%%%%%%%%%%%%%%%%%%%%%%%%%%%%%%%%%%%%%%%%%%%%%%%%%%%%%%%%%%%%%%

One can start analyzing the time evolution of the three classes of individuals. We numerically integrated Eqs. (\ref{eq1}), (\ref{eq2}) and (\ref{eq3}) in order to analyze the effects of the variation of the model's parameters. As initial conditions, we considered $s(0)=0.99$, $m(0)=0.01$ and $r(0)=0$, and for simplicity we fixed $\alpha=0.03$ and $\delta=0.07$, varying the parameters $\beta$ and $\gamma$. In Fig. \ref{fig1} (a), (b) and (c) we exhibit results for fixed $\beta=0.07$ and typical values of $\gamma$. One can see that the increase of $\gamma$ causes the increase of $s$ and $m$ and the decrease of $r$. Remembering that $\gamma$ models the persuasion of nonconsumers $S$ in the social interactions with risk drinkers $R$, i.e., the social pressure of individuals that do not consume alcohol over their contacts (friends, relatives, etc) that consume too much alcohol. On the other hand, in Fig. \ref{fig1} (d) we considered $\beta=0.20$ and $\gamma=0.15$. For this case, where we have $\beta>\gamma$, we see that the densities evolve in time, and in the steady states we observe the survival of only the risk drinkers, i.e., for $t\to\infty$ we have $s=m=0$ and $r=1$. This last result will be discussed in more details analytically in the following.

As we observed in Fig. \ref{fig1}, the densities $s(t)$, $m(t)$ and $r(t)$ evolve in time, and after some time they stabilize. In such steady states, the time derivatives of Eqs. (\ref{eq1}) - (\ref{eq3}) are zero. In the $t\to\infty$ limit, Eq. (\ref{eq1}) gives us $(-\beta\,m-\beta\,r+\gamma\,r)\,s=0$, where we denoted the stationary values as $s=s(t\to\infty)$, $m=m(t\to\infty)$ and $r=r(t\to\infty)$. This last equation has two solutions, one of them is $s=0$ and from the other solution we can obtain a relation between $r$ and $m$,
\begin{equation} \label{eq7}
r=\frac{\beta}{\gamma-\beta}\,m ~.
\end{equation}

Considering now the limit $t\to\infty$ in Eq. (\ref{eq2}), one obtains
\begin{equation} \label{eq8}
\beta\,s\,r=(\alpha+\delta\,r-\beta\,s)\,m ~.
\end{equation}
\noindent
If the obtained solution $s=0$ is valid, this relation gives us $m=0$ and consequently from (\ref{eq4}) we have $r=1$. This solution represents an absorbing state \cite{dos2013survival,de2012symbiotic}, since the dynamics becomes frozen due to the absence of $S$ and $M$ agents. We will discuss this solution in more details in the following.

Considering now the relation (\ref{eq7}) and the normalization condition (\ref{eq4}), one can obtain
\begin{equation} \label{eq9}
s=1-\frac{\gamma}{\gamma-\beta}\,m ~.
\end{equation}
\noindent
Substituting (\ref{eq9}) and (\ref{eq7}) in (\ref{eq8}) one obtains 2 solutions, $m=0$ and 
\begin{equation} \label{eq10}
m = \frac{[\gamma\,\beta-\alpha\,(\gamma-\beta)]\,(\gamma-\beta)}{\beta\,[\delta\,(\gamma-\beta)+\gamma^{2}]} ~.
\end{equation}
Considering this result (\ref{eq10}) in Eqs. (\ref{eq9}) and (\ref{eq7}) we obtain, respectively
\begin{eqnarray} \label{eq11}
s & = & 1 - \frac{\gamma\,[\gamma\,\beta-\alpha\,(\gamma-\beta)]}{\beta\,[\delta\,(\gamma-\beta)+\gamma^{2}]} ~, \\ \label{eq12}
r & = & \frac{\gamma\,\beta-\alpha\,(\gamma-\beta)}{\delta\,(\gamma-\beta)+\gamma^{2}} ~.
\end{eqnarray}
The obtained eqs. (\ref{eq10}) - (\ref{eq12}) represent a second possible steady state solution of the model, that is a realistic solution since the three fractions $s$, $m$ and $r$ coexist in the population.

We can look to Eq. (\ref{eq10}) in more details. It can be rewritten in the critical phenomena perspective as \cite{marro2005nonequilibrium,hinrichsen2000non}
\begin{equation} \label{eq13}
m = \frac{(\alpha+\gamma)\,(\beta-\beta_{c}^{(1)})\,(\beta_{c}^{(2)}-\beta)}{\beta\,[\delta\,(\gamma-\beta)+\gamma^{2}]} ~,
\end{equation}
\noindent
or in the standard form $m \sim (\beta-\beta_{c}^{(1)})\,(\beta_{c}^{(2)}-\beta)$, where $\beta_{c}^{(1)}=\alpha\,\gamma/(\alpha+\gamma)$ and $\beta_{c}^{(2)}=\gamma$. Thus, considering the density $m$ as a kind of order parameter, one observe in this model two distinct nonequilibrium phase transitions. The solution (\ref{eq10}) is valid in the range $\beta_{c}^{(1)}<\beta<\beta_{c}^{(2)}$. Notice that one can rewrite Eq. (\ref{eq12}) as  $r \sim (\beta-\beta_{c}^{(1)})$. In this case, one conclude that for $\beta<\beta_{c}^{(1)}$ the solutions (\ref{eq10}) - (\ref{eq12}) are not valid, since $m<0$ and $r<0$. Thus, in this region $\beta<\beta_{c}^{(1)}$ the valid solution is $m=r=0$ and from the normalization condition we have $s=1$. This last solution represents a second absorbing state, distinct from the first one obtained previously (where $s=m=0$ and $r=1$). Regarding this first absorbing state, it is valid in the region $\beta>\beta_{c}^{(2)}$.

\vspace{0.2cm}

Summarizing, the general solutions are:
\begin{align} \label{eq14}
s  =
\begin{cases} 1 & if \ \beta < \beta_{c}^{(1)}   
\\  
1-\frac{\gamma\,[\gamma\,\beta-\alpha\,(\gamma-\beta)]}{\beta\,[\delta\,(\gamma-\beta)+\gamma^{2}]} &
if \ \beta_{c}^{(1)}<\beta<\beta_{c}^{(2)}
\\ 
0 & if \ \beta > \beta_{c}^{(2)} 
\end{cases}
\end{align}

\begin{align} \label{eq15}
m  =
\begin{cases} 0 & if \ \beta < \beta_{c}^{(1)}   
\\  
\frac{[\gamma\,\beta-\alpha\,(\gamma-\beta)]\,(\gamma-\beta)}{\beta\,[\delta\,(\gamma-\beta)+\gamma^{2}]} & 
if \ \beta_{c}^{(1)}<\beta<\beta_{c}^{(2)}
\\ 
0 & if \ \beta > \beta_{c}^{(2)} 
\end{cases}
\end{align}

\begin{align} \label{eq16}
r  =
\begin{cases} 0 & if \ \beta < \beta_{c}^{(1)}   
\\  
\frac{\gamma\,\beta-\alpha\,(\gamma-\beta)}{\delta\,(\gamma-\beta)+\gamma^{2}} &
if \ \beta_{c}^{(1)}<\beta<\beta_{c}^{(2)}
\\ 
1 & if \ \beta > \beta_{c}^{(2)} 
\end{cases}
\end{align}
where the critical points are given by
\begin{eqnarray} \label{eq17}
\beta_{c}^{(1)} & = & \frac{\alpha\,\gamma}{\alpha+\gamma} ~, \\ \label{eq18}
\beta_{c}^{(2)} & = & \gamma ~.
\end{eqnarray}
\noindent

%%%%%%%%%%%%%%%%%%%%%%%%%%%%%%%%%%%%%%%%%%%%%%%%%%%%%%%%%%%%%%
\begin{figure}[t]
\begin{center}
\vspace{3mm}
\includegraphics[width=0.7\textwidth,angle=0]{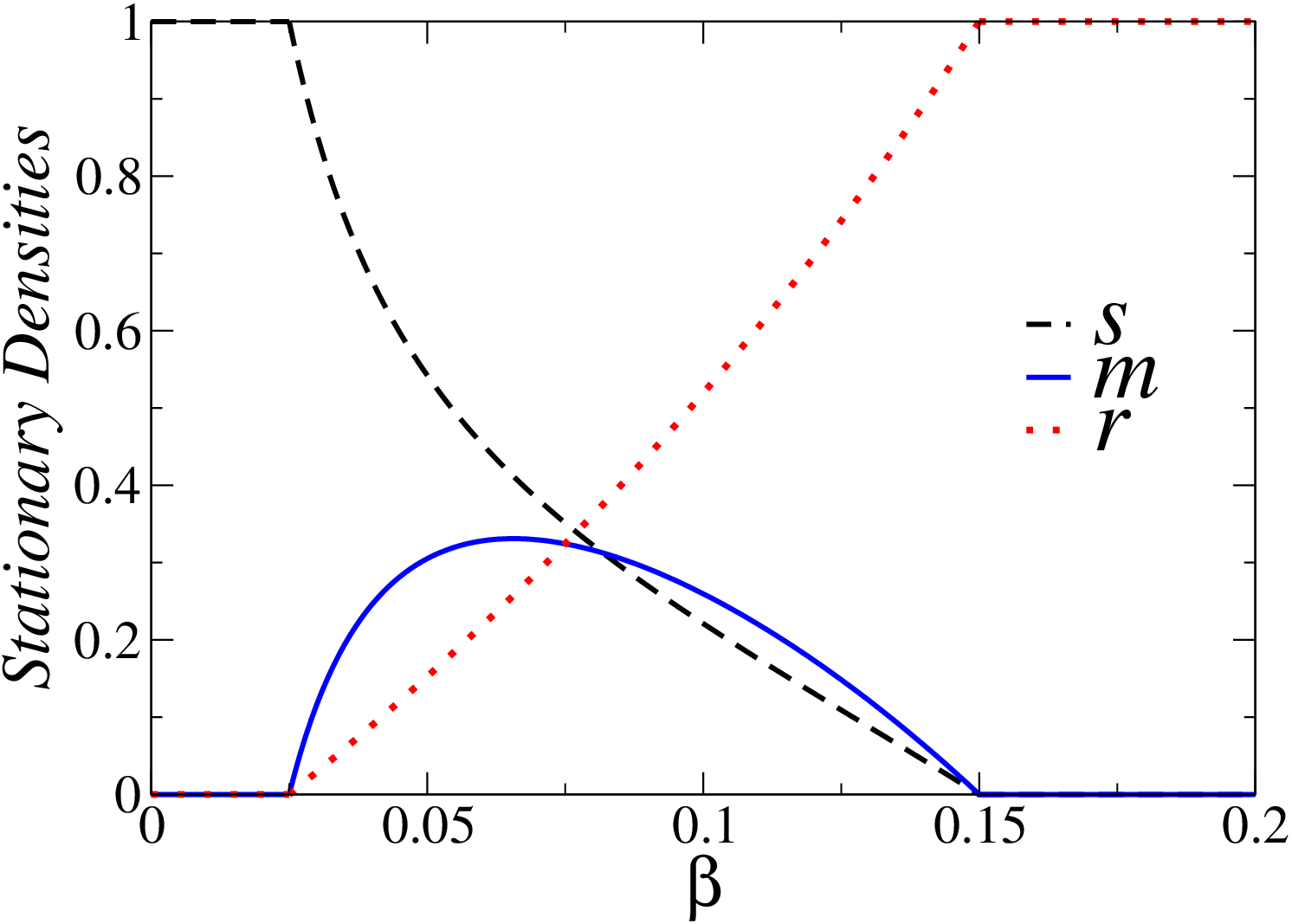}
\end{center}
\caption{(Color online) Stationary densities $s$, $m$ and $r$ as functions of the probability $\beta$ for $\alpha=0.03$, $\gamma=0.15$ and $\delta=0.07$. The lines were obtained from Eqs. \eqref{eq14} - \eqref{eq16}. For the considered parameters, the critical points are $\beta_{c}^{(1)}=0.025$ and $\beta_{c}^{(2)}=0.15$. It is important to note that these behaviors are present also for other parameter values, and what is shown here works as a pattern.}
\label{fig2}
\end{figure}
%%%%%%%%%%%%%%%%%%%%%%%%%%%%%%%%%%%%%%%%%%%%%%%%%%%%%%%%%%%%%%

Based on the results above, we plot in Fig. \ref{fig2} the steady state values of the three densities $s$, $m$ and $r$ as functions of $\beta$. For this graphic, we fixed the parameters $\alpha=0.03, \delta=0.07$ and $\gamma=0.15$. For such values, we have $\beta_{c}^{(1)}=0.025$ and $\beta_{c}^{(2)}=0.15$. As discussed previously, for $\beta<0.025$ the system is in one of the absorbing states in the long-time limit, where there are only nonconsumer agents in the population, i.e., $s=1$ and $m=r=0$. For $\beta>0.15$ the system becomes frozen in the other absorbing phase, where there are only risk drinkers in the population after a long time, i.e., $r=1$ and $s=m=0$. Among those states, we have a realistic region where all the three kinds of individuals, nonconsumers, moderate drinkers and risk drinkers coexist in the population.

The competition among the contagions cause the occurrence of such three regions in the model. From one side we have drinkers (moderate and risk) influencing nonconsumers to consume alcohol, with probability $\beta$. On the other hand, we have the social pressure of nonconsumers over risk drinkers, with probability $\gamma$, in order to make such alcoholics to begin treatment and stop drinking. Finally, it is important to mention the parameter $\alpha$, that drives the only transition of the model that does not depend on a direct social interaction. That parameter models the spontaneous increase of alcohol consumption, and it is also responsible for the first phase transition (together with $\gamma$), since we have $\beta_{c}^{(1)}=0$ for $\alpha=0$. It means that the alcohol consumption (the "disease") cannot be eliminated of the population after a long time if there is a spontaneous increase of alcohol consumption from individuals that drink moderately, which is a realistic feature of the model.

%%%%%%%%%%%%%%%%%%%%%%%%%%%%%%%%%%%%%%%%%%%%%%%%%%%%%%%%%%%%%%
\begin{figure}[t]
\begin{center}
\vspace{3mm}
\includegraphics[width=0.7\textwidth,angle=0]{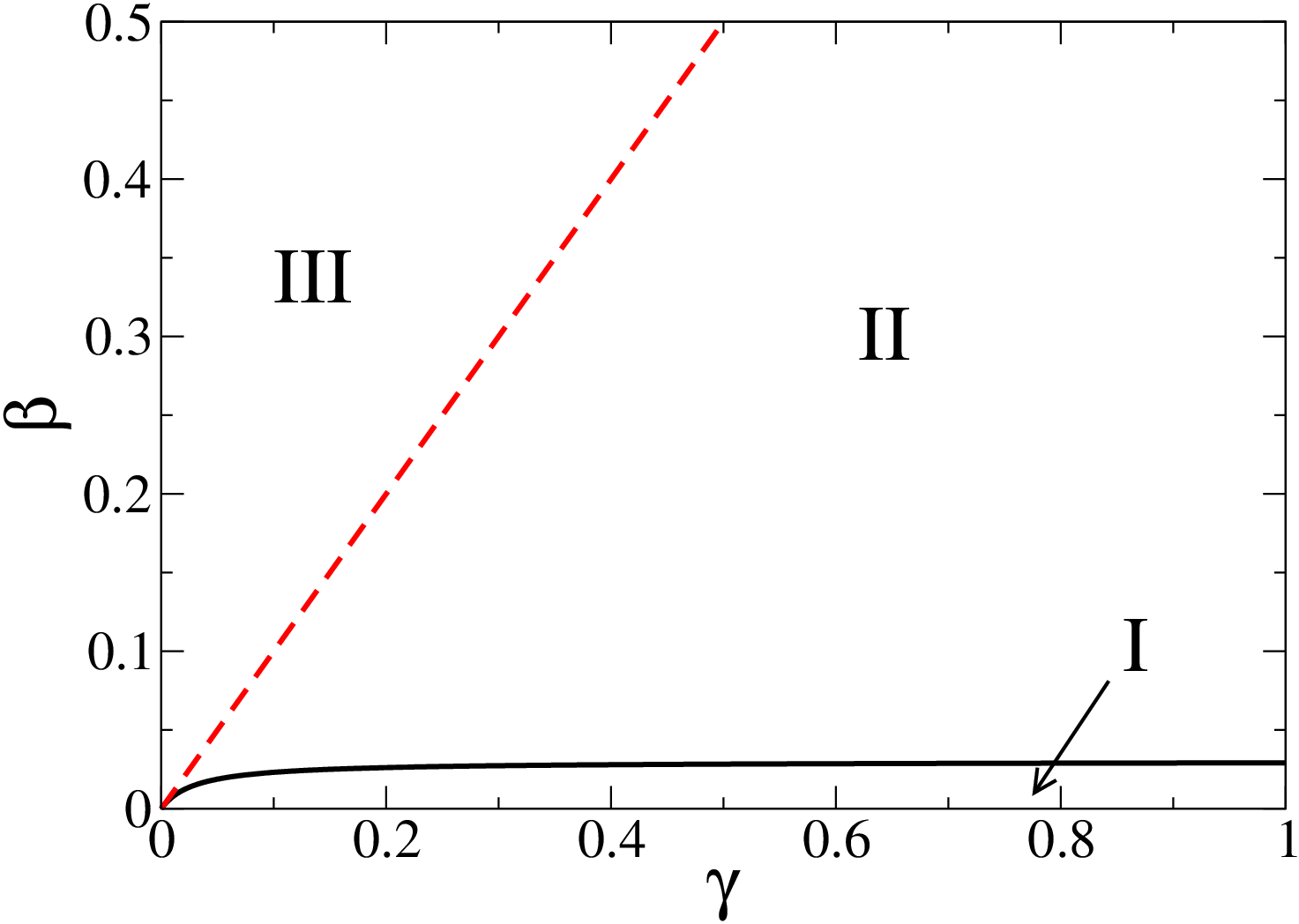}
\end{center}
\caption{(Color online) Phase diagram of the model in the plane $\beta$ \textit{vs} $\gamma$ for $\delta=0.07$ and $\alpha=0.03$. The full line represents the critical point $\beta_{c}^{(1)}$ given by Eq. (\ref{eq17}) and the dashed line the other critical point $\beta_{c}^{(2)}$ given by Eq. (\ref{eq18}). \textbf{I} denotes the region where the system falls in the absorbing phase $s=1, m=r=0$, \textbf{II} denotes the region where the three densities $s$, $m$ and $r$ coexist and \textbf{III} denotes the region where the system falls in the other absorbing phase $s=m=0, r=1$.}
\label{fig3}
\end{figure}
%%%%%%%%%%%%%%%%%%%%%%%%%%%%%%%%%%%%%%%%%%%%%%%%%%%%%%%%%%%%%%

For clarity, we exhibit in Fig. \ref{fig3} the phase diagram of the model in the plane $\beta$ versus $\gamma$, separating the three above discussed regions. In Fig. \ref{fig3}, the absorbing phase with $s=1$ and $m=r=0$ is located in region I for $\beta<\beta_{c}^{(1)}$, the coexistence phase is denoted by II for $\beta_{c}^{(1)}<\beta<\beta_{c}^{(2)}$ (where the three densities coexist) and the other absorbing phase where $s=m=0$ and $r=1$ is located in region III. From this figure we see the mentioned competition among the contagions. Indeed, if $\beta$ is sufficiently high, many nonconsumers become moderate drinkers. Such moderate drinkers will become risk drinkers (via probabilities $\alpha$ and $\delta$), and in the case of small $\gamma$ we will observe after a long time the disappearance of nonconsumers and moderate drinkers (region III). In the opposite case, i.e., for high $\gamma$ and small $\beta$, the flux into the compartment S is intense, and in the long-time limit the other two subpopulations M and R disappears (region I). Finally, for intermediate values of $\beta$ and $\gamma$ the competition among the social interactions lead to the coexistence of the three subpopulations in the stationary states (region II). It is worthwhile to mention that the sizes of regions I and II are directly dependent on probability $\alpha$, while region III is always fixed due to Eq. (\ref{eq18}). This means that, if parameter $\alpha$ is increased, region I will become gradually larger, which is an indication that the spontaneous evolution from moderate to risk drinking behavior increases the latter's absorbing state. In consequence, since probability $\alpha$ represents a percentage of moderate drinkers that become risk drinkers without the need for social interaction, it is a crucial factor not only to implement the theoretical model but also to identify a possible percentage of the population that has a natural tendency to present excessive alcohol consumption behavior, regardless of their social interaction network. For Fig. \ref{fig3}, for instance, this value is $3\%$. Larger values of $\alpha$ narrow the set of parameters that can be chosen in order to realistically describe a real system.

%%%%%%%%%%%%%%%%%%%%%%%%%%%%%%%%%%%%%%%%%%%%%%%%%%%%%%%%%%%%%%
\begin{figure}[t]
\begin{center}
\vspace{3mm}
\includegraphics[width=0.7\textwidth,angle=0]{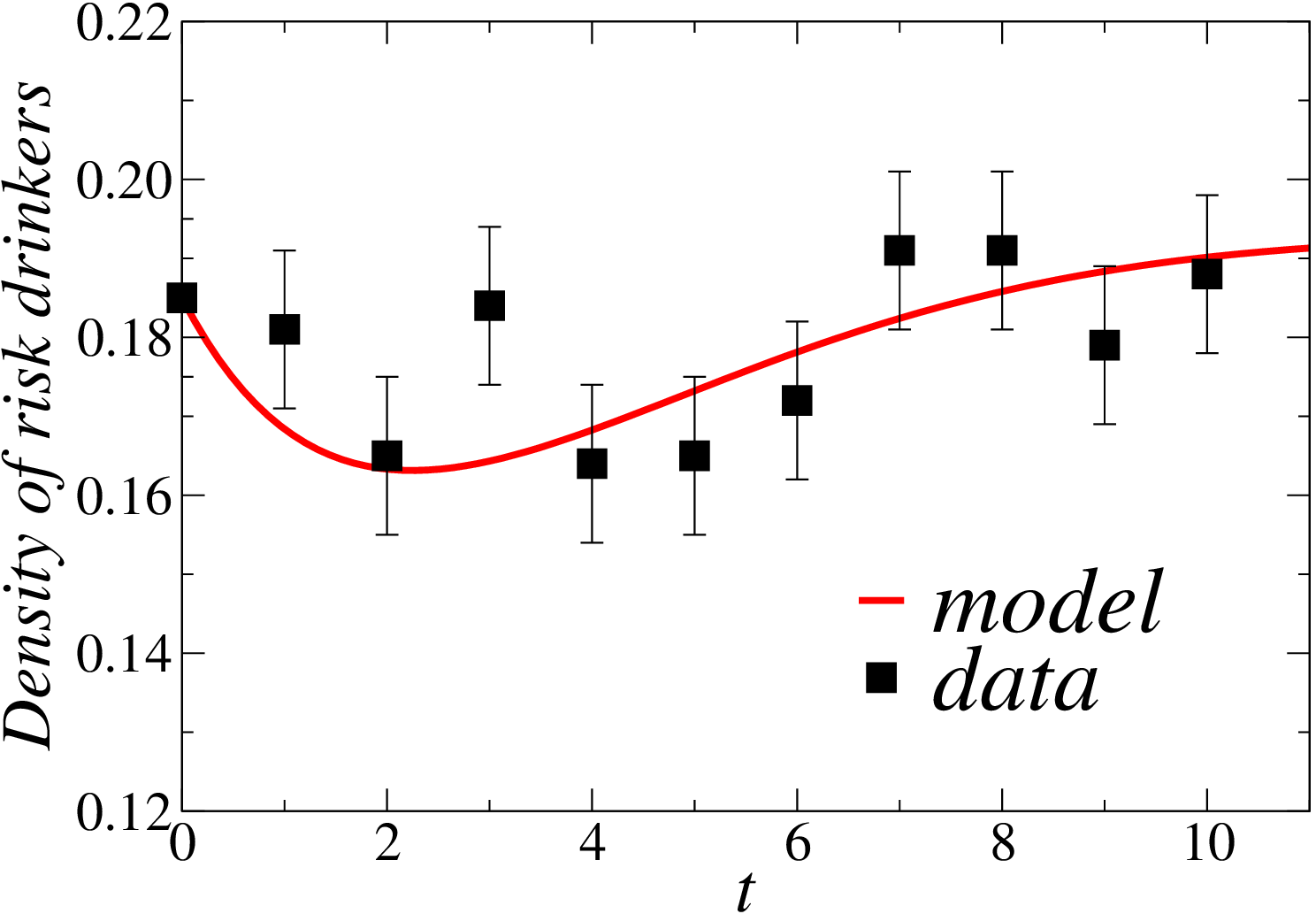}
\end{center}
\caption{(Color online) Comparison between data of abusive alcohol consumption in Brazil through time and the time evolution of the density of risk drinkers $r(t)$ given by the numerical integration of Eqs. (\ref{eq1}) - (\ref{eq3}). Parameters are $\beta=0.06$, $\gamma=0.11$, $\alpha=0.047$ and $\delta=0.2$.}
\label{fig4}
\end{figure}
%%%%%%%%%%%%%%%%%%%%%%%%%%%%%%%%%%%%%%%%%%%%%%%%%%%%%%%%%%%%%%

Finally, in order to verify if this simple three-compartment model is able to qualitatively reproduce real alcohol consumption data, we compare the model's results with official data of drinking consumption in Brazil. In reference \cite{databrasil} there is data only for the abusive alcohol consumption. Data were collected from 2009 to 2019, thus in Fig. \ref{fig4} the initial time $t=0$ represents the fraction of abusive drinkers for 2009, $t=1$ represents the fraction for 2010, and so on. Since the data is only for the fraction of people that consume alcohol abusively, we plot the density of risk drinkers $r(t)$ together with the data. In order to compare them with the model, we considered for the initial density of risk drinkers $r(0)=0.185$ and numerically integrated Eqs. (\ref{eq1}) - (\ref{eq3}). The value $0.185$ was chosen since it is the fraction of abusive drinkers for 2009 obtained from the database \cite{databrasil}. In addition, we re-scaled the time of the simulation results to match the time of real data: such simulation time was multiplied by $0.12$ for a better comparison. We find that the simulated drinking trajectories qualitatively correspond to the data. For the numerical results, we considered the parameters  $\beta=0.06$, $\gamma=0.11$, $\alpha=0.047$ and $\delta=0.2$, which indicate that the probability of finding an individual that will spontaneously become a risk drinker in Brazil during the last decade is around 4.7\%. Furthermore, looking at Eqs. (\ref{eq17}) and (\ref{eq18}), it is easy to see that in order to model Brazil's population we have $\beta_{c}^{(1)} = 0.033$ and $\beta_{c}^{(2)} = 0.11$, which is in accordance with the relation $\beta_{c}^{(1)}<\beta<\beta_{c}^{(2)}$, showing that the model describes the available data in its most realistic spectrum (region II of Fig. \ref{fig3}). Naturally, in comparison with actual data, models should always present the three different phases at its equilibrium phase, i.e. coexistence between the three different population groups, since descriptions tending to steady states with only nonconsumers or risk drinkers are unrealistic. This qualitative agreement with Brazil's database in the realistic spectrum of the model points to a good, albeit simplistic, modeling.

% ###############################################################

\section{Final remarks}   

In this work, we have studied a compartmental model that aims to describe the evolution of drinking behavior in an adult population. We considered a fully-connected population that is divided in three compartments, namely Susceptible individuals S (nonconsumers), Moderate drinkers M and Risk drinkers R. The transitions among the compartments are ruled by probabilities, representing the social interactions among individuals, as well as spontaneous decisions, in particular from moderate evolving into risk drinkers, and we studied the model through analytical and numerical calculations.

From the theoretical point of view, the model is of interest of Statistical Physics since we observed the occurrence of two distinct nonequilibrium phase transitions. These transitions separate the model in three regions: (I) existence of nonconsumers only; (II) coexistence of the three compartments and (III) existence of risk drinkers only. Regions I and III represent two distinct absorbing phases, since the system becomes frozen due to the existence of only one subpopulation for each case - this means that, in order to describe real populational systems, the parameters must be chosen so that the model falls in region II, since populations consisting solely of nonconsumers or risk drinkers do not represent a realistic entity. The critical points of such transitions were obtained analytically. 

A comparison with available data for Brazil's extreme alcohol consumption for the past decade shows a good qualitative agreement with the model, with the chosen parameters framed within its realistic boundaries. It is interesting to note that a simple model like the one presented here is able, with only four parameters, to qualitatively describe the evolution for alcohol consumption. This shows that the dynamic parameters chosen to describe the transitions between the three different compartments can be good candidates to model realistic behavior when dividing the population into these three groups. It will be important in a couple of years time to re-evaluate these results in the light of new data comprising years 2020 and 2021, in order to verify the direct effects of the COVID-19 pandemic in the Brazilian population's alcohol consumption. An hypothesis to be tested is a possible increase in parameter $\alpha$ combined with a corresponding decrease in the other parameters, corresponding to social interactions.

As future extensions, it can be considered the inclusion of heterogeneities in the population, like agents' conviction \cite{crokidakis2012competition}, time-dependent transition rates \cite{crokidakis2012critical}, inflexibility \cite{crokidakis2015inflexibility,galam2007role}, mass media effects \cite{huo2017optimal,muthuri2019modeling}, social temperature \cite{nuno_celia2017} etc.

\section*{Acknowledgments}

The authors thank Ronald Dickman for some suggestions. Financial support from the Brazilian scientific funding agencies CNPq (Grants 303025/2017-4 and 311019/2017-0) and FAPERJ (Grant 203.217/2017) is also acknowledge.

%The authors acknowledge financial support from the Brazilian scientific funding agencies CNPq and FAPERJ. 



%% References



%% New version of the num-names style
\bibliographystyle{elsarticle-num-names}

%\bibliography{alcoholism.bib}

\begin{thebibliography}{00}

\bibitem{bailey1975mathematical}
N. T. Bailey, \textit{The mathematical theory of infectious diseases and its applications} (Charles Griffin \& Company Ltd, 5a Crendon Street, High Wycombe, Bucks HP13 6LE, 1975).

\bibitem{daley1964epidemics}
D. J. Dailey, D. G. Kendall, Epidemics and Rumours, Nature 204 (1964) 1118-1118.

\bibitem{PMID:21696936}
F. Guerrero, F.-J. Santonja, R.-J. Villanueva, Analysing the Spanish smoke-free legislation of 2006: a new method to quantify its impact using a dynamic model, The International Journal on Drug Policy 22 (2011) 247-251.


\bibitem{sanchez2011predicting}
E. S{\'a}nchez, R.-J. Villanueva, F.-J. Santonja, M. Rubio, Predicting cocaine consumption in Spain: A mathematical modelling approach, Drugs: Education, Prevention and Policy 18 (2011) 108-115.

\bibitem{santonja2010alcohol}
F.-J. Santonja, E. S\'anchez, M. Rubio, J.-L. Morera,  Alcohol consumption in Spain and its economic cost: a mathematical modeling approach, Mathematical and Computer Modelling 52 (2010) 999-1003.

\bibitem{ejima2013modeling}
K. Ejima, K. Aihara, H. Nishiura, Modeling the obesity epidemic: social contagion and its implications for control, Theoretical Biology and Medical Modelling 10 (2013) 17.


\bibitem{crokidakis2018can}
N. Crokidakis, J. S. Sa Martins, Can honesty survive in a corrupt parliament?, International Journal of Modern Physics C 29 (2018) 1850094.

\bibitem{lima2014evolution}
F. Lima, T. Hadzibeganovic, D. Stauffer, Evolution of tag-based cooperation on Erd{\H{o}}s-R\'enyi random graphs, International Journal of Modern Physics C 25 (2014) 1450006.

\bibitem{marvel2012encouraging}
S. A. Marvel, H. Hong, A. Papush, S. H. Strogatz, Encouraging moderation: clues from a simple model of ideological conflict, Physical Review Letters 109 (2012) 118702.

\bibitem{gunduz2016dynamics}
G. G{\"u}nd{\"u}z, The dynamics of the rise and fall of empires,  International Journal of Modern Physics C 27 (2016) 1650123.

\bibitem{brum2017dynamics}
R.  M.  Brum,  N.  Crokidakis, Dynamics  of  tax  evasion  through  an epidemic-like model, International Journal of Modern Physics C 28 (2017) 1750023.


\bibitem{galam2016modeling}
S. Galam, M. A. Javarone, Modeling radicalization phenomena in heterogeneous populations, PloS one 11 (2016) e0155407.

\bibitem{stauffer2007can}
D. Stauffer, M. Sahimi, Can a few fanatics influence the opinion of a large  segment  of  a  society?, The European Physical Journal B 57 (2007) 147-152.

\bibitem{nizamani2014public}
S. Nizamani, N. Memon, S. Galam, From public  outrage  to  the  burst  of  public  violence:  An epidemic-like  model, Physica A:  Statistical  Mechanics  and  its  Applications 416 (2014) 620-630.

\bibitem{galea2009social}
S.  Galea,  C.  Hall,  G.  A.  Kaplan,  Social epidemiology  and complex system dynamic modelling as applied to health behaviour and drug use research, International Journal of Drug Policy 20 (2009) 209-216.

\bibitem{gorman2006agent}
D.  M.  Gorman,  J.  Mezic,  I.  Mezic,  P. J. Gruenewald,   Agent-based modeling of drinking behavior: a preliminary model and potential applications to theory and practice,  American Journal of Public Health 96 (2006) 2055-2060.

\bibitem{oms2018}
WHO, World Health Statistics 2018: Monitoring  health  for  the  sdgs, sustainable development goals, Geneva:  World Health Organization CCBY-NC-SA (2018) 3.0 IGO.

\bibitem{sullivan2005}
L.  E.  Sullivan,  D.  A.  Fiellin,  P.  G.  O'Connor, The  prevalence  and impact  of alcohol  problems in  major  depression:  a  systematic  review, The American Journal of Medicine 118 (2005) 330-341.

\bibitem{chodkiewicz2020alcohol}
J. Chodkiewicz, M. Talarowska, J. Miniszewska, N. Nawrocka, P. Bilinski,  Alcohol consumption reported during the COVID-19 pandemic: The initial stage, International Journal of Environmental Research and Public Health 17 (2020) 4677.

\bibitem{clay2020alcohol}
J. M. Clay, M. O. Parker,  Alcohol use and misuse during the COVID-19 pandemic: a potential public health crisis?, The Lancet Public Health 5 (2020) e259.

\bibitem{narasimha2020complicated}
V. L. Narasimha, L. Shukla, D. Mukherjee, J. Menon, S. Huddar, U. K. Panda, J. Mahadevan, A. Kandasamy, P. K. Chand, V. Benegal, P. Murthy,Complicated alcohol withdrawal - An unintended consequence of COVID-19 lockdown, Alcohol and Alcoholism (Oxford, Oxfordshire) (2020).

\bibitem{rehm2020alcohol}
J. Rehm, C. Kilian, C. Ferreira-Borges, D. Jernigan, M. Monteiro, C. D. Parry,  Z.  M.  Sanchez,  J.  Manthey,   Alcohol  use  in  times  of  the  COVID-19:  Implications for monitoring and policy, Drug and Alcohol Review 39 (2020) 301-304.

\bibitem{walters2013modelling}
C. E. Walters, B. Straughan, J. R. Kendal, Modelling alcohol problems: total recovery, Ricerche di Matematica 62 (2013) 33-53.

\bibitem{nazir2019conformable}
A. Nazir, N. Ahmed, U. Khan, S. T. Mohyud-Din, A conformable mathematical model for alcohol consumption in Spain,  International Journal of Biomathematics 12 (2019) 1950057.

\bibitem{huo2018dynamics}
H.-F. Huo, H.-N. Xue, H. Xiang,  Dynamics of an alcoholism model on complex  networks  with  community  structure  and  voluntary  drinking, Physica A: Statistical Mechanics and its Applications 505 (2018) 880-890.

\bibitem{mulone2012modeling}
G. Mulone, B. Straughan, Modeling binge drinking, International Journal of Biomathematics 5 (2012) 1250005.

\bibitem{sanchez2007drinking}
F. S\'anchez, X. Wang, C. Castillo-Ch\'avez, D. M. Gorman, P. J. Gruenewald,  Drinking as an epidemic- a simple mathematical model with recovery and relapse, in: Therapist’s Guide to Evidence-Based Relapse Prevention, Elsevier (2007) 353-368.

\bibitem{sharma2013drinking}
S. Sharma, G. P. Samanta, Drinking as an Epidemic: A mathematical model with dynamic behaviour,  Journal of Applied Mathematics \& Informatics 31 (2013) 1-25.

\bibitem{agrawal2018role}
A. Agrawal, A. Tenguria, G. Modi, Role of epidemic model  to  control  drinking  problem, International Journal of Scientific Research in Mathematical and Statistical Sciences 5(4) (2018) 324-337.

\bibitem{adu2017mathematical}
I. K. Adu, M. AL-Rahman EL-Nor Osman, C. Yang, Mathematical model of drinking epidemic, Journal of Advances in Mathematics and Computer Science, 22(5) (2017) 1-10. https://doi.org/10.9734/BJMCS/2017/33659.

\bibitem{huo2012global}
Hai-Feng Huo, Na-Na Song, Global stability for a binge drinking model with two stages, Discrete Dynamics in Nature and Society, vol. 2012, Article ID 829386, 15 pages, 2012. https://doi.org/10.1155/2012/829386.

\bibitem{khajji2020discrete}
B. Khajji, A. Labzai, A. Kouidere, O. Balatif,  M. Rachik, A discrete mathematical modeling of the influence of alcohol treatment centers on the drinking dynamics  using  optimal control,  Journal of Applied Mathematics, vol. 2020, Article ID 9284698, 13 pages, 2020. https://doi.org/10.1155/2020/9284698.

\bibitem{huo2017optimal}
Hai-Fen Huo, Shui-Rong Huang, Xun-Yang Wang, H. Xiang, Optimal control of a social epidemic model with media coverage, Journal of Biological Dynamics 11 (2017) 226-243.


\bibitem{muthuri2019modeling}
G. G. Muthuri, D. M. Malonza, F. Nyabadza,  Modeling the effects of treatment on alcohol abuse in Kenya incorporating mass media campaign, Journal of Mathematical and Computational Science 9 (2019) 632-653.

\bibitem{ma2015modelling}
Shuang-Hong Ma, Hai-Feng Huo, Xin-You Meng, Modelling alcoholism as a contagious disease: A mathematical model with awareness programs and time delay, Discrete Dynamics in Nature and Society 2015 (2015).

\bibitem{wang2014optimal}
Xun-Yang Wang, Hai-Feng Huo, Qing-Kai Kong, Wei-Xuan Shi, Optimal control strategies in an alcoholism model, in:  Abstract and Applied Analysis, volume 2014, Hindawi, 2014.


\bibitem{databrasil}
Brazil's Health Ministry, Vigitel Brasil 2019 : vigil\^ancia de fatores de risco e proteção para doenças crônicas por inquérito telefônico : estimativas sobre frequência e distribuição sociodemográfica de fatores de risco e proteção para doenças crônicas nas capitais dos 26 estados brasileiros e no Distrito Federal em 2019, $https://bvsms.saude.gov.br/bvs/publicacoes/vigitel\_brasil\_2019\_vigilancia\_fatores\_risco.pdf$ (2020).


\bibitem{crokidakis2020covid}
N. Crokidakis,  COVID-19  spreading  in  Rio  de  Janeiro,  Brazil: Do  the policies of social isolation really work?,  Chaos, Solitons \& Fractals 136 (2020) 109930.

\bibitem{dos2013survival}
R. V. dos Santos, R. Dickman, Survival of the scarcer in space, Journal of Statistical Mechanics:  Theory and Experiment 2013 (2013) P07004.

\bibitem{de2012symbiotic}
M. M. de Oliveira, R. V. dos Santos, R. Dickman, Symbiotic two-species contact process, Physical Review E 86 (2012) 011121.


\bibitem{marro2005nonequilibrium}
J. Marro, R. Dickman, \textit{Nonequilibrium phase transitions in lattice models} (Cambridge University Press, 2005).


\bibitem{hinrichsen2000non}
H.  Hinrichsen,  Non-equilibrium  critical  phenomena  and  phase  transitions into absorbing states,  Advances in Physics 49 (2000) 815-958.


\bibitem{crokidakis2012competition}
N. Crokidakis, F. L. Forgerini, Competition among reputations in the 2d Sznajd model:  Spontaneous emergence of democratic states,  Brazilian Journal of Physics 42 (2012) 125-131.

\bibitem{crokidakis2012critical}
N. Crokidakis, M. A. de Menezes,  Critical behavior of the SIS epidemic model  with  time-dependent  infection  rate,  Journal  of  Statistical Mechanics: Theory and Experiment 2012 (2012) P05012.

\bibitem{crokidakis2015inflexibility}
N.  Crokidakis,  P.  M.  C.  de  Oliveira,   Inflexibility  and  independence: Phase  transitions  in  the  majority-rule  model,   Physical Review  E  92 (2015) 062122.

\bibitem{galam2007role}
S. Galam,  F. Jacobs,  The  role  of  inflexible  minorities  in the  breaking  of  democratic  opinion  dynamics,   Physica  A:  Statistical Mechanics and its Applications 381 (2007) 366-376.

\bibitem{nuno_celia2017}
C. Antenodo, N. Crokidakis, Symmetry breaking by heating in a continuous opinion model ,Physical Review E 95 (2017) 042308.


\end{thebibliography}






% \end{thebibliography}

\end{document}